## Generalized kinetics of overall phase transition in terms of logistic equation

I. Avramov, J. Šesták *#

Institute of Physical Chemistry, the Bulgarian Academy of Sciences, Bg - 1113 Sofia, Bulgaria, E-mail: avramov@ipc.bas.bg

* New Technologies - Research Centre of the Westbohemian Region (NTC) of the University of West Bohemia, Universitní 8, CZ - 301 14 Plzeň

# Section of Solid-State Physics of Institute of Physics, the Academy of Sciences of Czech Republic, Cukrovarnicka 10, CZ-16200 Praha, both Czech Republic; E-mail: sestak@fzu.cz

E-mail for correspondence: avramov@ipc.bas.bg

### Abstract

We summarize and to discuss briefly the geometrical practice of modeling attitudes so far popular in treating reaction kinetics of solid-state processes. The model equations existing in the literature have been explored to describe the thermal decomposition and crystallization data and are deeply questioned and analyzed showing that under such a simple algebraic representation, the reacting system is thus classified as a set of geometrical bodies (~ spheres) where each and every one reaction interface is represented by similar and smooth characteristics of reaction curve. It brings an unsolved question whether the sharp and even boundary factually exists or if it resides jointly just inside the global whole of the sample entirety preventing individual particles from having their individual reaction front. Most of the derived expressions are specified in an averaged generalization in terms of the three and two parameters equation (so called *JMAK* and *SB* models) characterized by a combination of power exponents *m*, *n* and *p* as summarized in a lucid Table. As an alternative the logistic equation is proposed powered with fractal exponents standing for the interfaces to be identified with an underlying principle of *defects.* Unfortunately, many of the solutions for the standard kinetic equations are truncated by infinite series, unfriendly to mathematical solutions. Based on the assumption that transformation rate is a product of two functions $f(\alpha)k(t)$, we propose a fundamentally new method to analyze the individual mechanism of each

process.The idea is to plot the experimental data in coordinates *the transformation rate* against *f(α)*.

**Keywords:** logistic function, phase transition, crystallization, overall process, Avrami, *JMAK* and *SB* equations, fractal power exponents

## Introduction

In the turn of Forties, Avrami [1-3] published a fundamental paper on the pages of this Journal instigating a geometrical concept for the description of reaction mechanism occurring in the solid-state processes. Since that, similar algebraic portrayal, based on well defined geometrical bodies[4], became popular in thermal analysis kinetics [5-11] overshadowing other approaches for its apparent simplicity and straightforward visual imagination. Such a geometrical manner of kinetic modeling was multiple reiterated [4,12,13] still exploiting, however, a strict vision of ideal Euclidian bodies. This image was applied to various reaction models the most common being the simple shrinking-core model based on a globular particle [4-13] which, however, necessitates a sharp reaction boundary. Upon a simple geometrical representation, the reacting system is thus classified as a set of spheres where each and every one reaction interface is represented by similar characteristics of reaction curve. It brings an unsolved question whether the sharp boundary factually exists within the each particle assuming homogeneous temperature distribution or if it resides jointly just inside the global whole of the sample entirety preventing individual particles from having their individual reaction front[10]. Moreover such a modeling is responsible to depict the incorporated particles within the reaction interfaces by smooth *disjointing lines* the evenness of which does not match up the surface reality of multi-particle samples seen, e.g., by microscopy. The derived kinetics then depends on the behavior of such interfaces separating the product and the initial reactant. Accordingly the space co-ordinates of the rate-controlling elements become a design respecting heterogeneity.

Nevertheless there are multiple papers using the above mentioned system of kinetic analysis for a successful describing of reaction mechanisms, e.g. recent [14-23], which, however, frequently accept the emergency of non-integral dimensionalities – fractal constants. It means that interfaces produced during a reaction can be identified with an underlying principle of *defects* symbolized by a pictographic contour (borderline curves) at our graphical portrayal. While traditionally identifying reaction mechanism with such an assemble of separated particles and consequently exploring the sample's overall behavior under a real course of measurement based onthe observation of its averaged properties we may alternatively see the studied process in terms of a global propagation of ′defects′ throughout the sample. Such a collective way of the investigation of reaction path may thus better correlate to the spread-out logistic-like philosophy under a simplified model-free description. It is enriched by using a ‘blank’ pattern within the double exponents’ framework of a *logistic equation* making possible to interconnect it with standard structure of models. .

**Logistic functions**

In 1844 Belgium professor Pierre Francois Verhulst (1804-1849) gave the name logistic functions to sigmoidal expressions[24-26] describing the expected population growth in his country. Generalized logistic curves offer models to the "S-shaped" behavior of creation of new state. The initial stage of growth is approximately exponential; then, as saturation, the growth slows and stops. There are many processes the time dependencies of which can be described by logistic curves [27-32]. In addition to the field of materials science similar behavior was found for the pestilence spread among individuals, the propagation of population of plants, or of microorganisms, and many others cases. Actually, all mentioned processes can be described by variants of logistic functions[25–32] because they describe the evolution in an environment with a bordered upper limit of the population, describable by dimensionless degree of conversion, α. It is natural to assume[11,12,33,34] that the transformation rate $\frac{d\alpha}{dt}$ is a product of two functions

$$\frac{d\alpha}{dt} = f(\alpha)k(t) \tag{1}$$

The first one, *f(α)*, accounts for the so-called reaction mechanism being explicitly dependent on the transformation progress, *α*; the other one, *k(t)*, depending eventually, on time, *t*. The main idea of this assumption is to separate variables in order to solve the differential equation:

$$F(\alpha) \equiv \int_0^\alpha \frac{d\alpha'}{f(\alpha')} = \int_0^t k(t)dt \equiv K(t) \tag{2}$$

First, we discuss two frequently used approximations for the *k(t)* function, respectively for the solutions of the right-hand integral in Eq.(2). If the system is kept under constant conditions, it is also logical to assume that $k(t) = k$ is constant so that

$$K(t) = \int_0^t k(t)dt = kt\,. \tag{3}$$

The second, frequently used solution, concerns the case when the system is heated (or cooled) under constant rate of heating, $q = \frac{dT}{dt}$. Under these conditions the time dependence of temperature is $T = T_{in} + qt$. The time integral transforms to $\int_0^t k(t)dt = \frac{1}{q}\int_0^T k(T)dT$. Here the initial temperature $T_{in}$ is replaced by zero because at low temperatures the transformation rate is negligibly slow. If the characteristic time is determined by an activation energy, *E,* according to expression

$$k(T) = \frac{1}{\tau_o}\exp\left[-\frac{E}{RT}\right], \qquad (4)$$

the approximate solution *K(T)* of the corresponding integral (see [37]) is

$$K(T) = \frac{1}{q}\int_0^T k(T)dT \approx \frac{RT^2}{q\tau_o E}\exp\left(-\frac{E}{RT}\right). \qquad (5)$$

If there are enough reasons to believe that the conditions are not essentially changing during the process, i.e., if τ can be assumed constant, it is appropriate to replace in the table *K(t)* by t/τ as follows from Eq.(3) (for constant conditions), or *K(T),* determined by Eq.(5) (for constant heating/cooling) rates).

A special case was suggested by Sestak and Berggren [34-36] recommending a mathematical form that represents a generalized *f(α)* function:

$$f(\alpha) = \alpha^m (1-\alpha)^n (-\ln(1-\alpha))^p \qquad (6)$$

where *m, n,* and *p* are constants. By assigning values for these three variables, any classically geometrical model can be expressed; however, the pair combination of *m* and *n* is the most reliable[36] thus often called the *SB-equation*[35,36]. In Eq.(2) the terms $\alpha^m$ *and* $(-\ln(1-\alpha))^p$ reflect that the nature of process is autocatalytic, i.e., the transformation rate increases with α. On the other hand, the term $(1-\alpha)^n$ accounts that the system cannot support more than $\alpha$=1 parts in the new state.
Derivations and theoretical implications of the values of *m, n,* and *p* are discussed below concerning specific models.

We discuss the reasons to make assumptions on the values of m, n, and p, i.e., the form of *f(α)* function, and to what kind of complex systems each of the proposed assumption stays reasonably enough. The results for various models routinely treated in literature[4,12,13-22] are summarized in Table I.

*Table I:* ***Expressions for individual cases of mathematical simulation***

| **Exponents = n, m, p** | ***F(α)*** | ***Solution*** | ***Remarks*** |
|---|---|---|---|
| n = 1; m= 1, p=0<br>$f(\alpha)=\alpha(1-\alpha)$ | $\ln\frac{\alpha(1-\alpha_o)}{\alpha_o(1-\alpha)}$ | $\alpha(t)=\frac{\alpha_o}{\alpha_o+\exp(-K(t))}$ | Classical logistic function (Verhulst)<br>Autocatalytic reactions (Prout-Tompkins)<br>Autogenesis Akulov |
| n =0; m= random (≠1), p=0<br>$f(\alpha)=\alpha^m$ | $\frac{\alpha_o^{1-m}}{1-m}-\frac{\alpha^{1-m}}{1-m};m\neq 1$ | *Unlimited growth*<br>$\alpha(t)=\left(\frac{1}{\left(\frac{1}{\alpha_o}\right)^{m-1}-(m-1)K(t)}\right)^{\frac{1}{m-1}}$<br>for<br>$0<K(t)<\frac{\left(\frac{1}{\alpha_o}\right)^{m-1}-1}{(m-1)}$ | Kopelman 1986<br>Fractal variant |
| n =0; m= 1, p=0<br>$f(\alpha)=\alpha$ | $\ln\frac{\alpha}{\alpha_o}\quad if\ \ m=1$ | $\alpha=\alpha_o\exp(K(t))$ for<br>$0\le K(t)\le-\ln\alpha_o$ | classical homogeneous reaction |
| $f(\alpha)=\alpha^m$<br>n =0; **m= 2**, p=0<br><br>**m= 3** | $\frac{1}{\alpha_o}-\frac{1}{\alpha}\ \ if\ \ m=2$<br><br>$\frac{1}{2\alpha_o^2}-\frac{1}{2\alpha^2};m=3$ | $\alpha(t)=\frac{1}{\left(\frac{1}{\alpha_o}\right)-K(t)};for\ m=2$<br>$0\le K(t)\le\frac{1}{\alpha_o}-1$<br>$\alpha(t)=\sqrt{\frac{1}{\left(\frac{1}{\alpha_o}\right)^2-2K(t)}}\ for\ m=3$<br>$0\le K(t)\le\frac{\left(\frac{1}{\alpha_o}\right)^2-1}{2}\approx\frac{1}{2\alpha_o^2}$ | classical homogeneous reaction<br>Phase boundary reactions<br>Contracting |
| n = random; m=0, p=0 $f(\alpha)=(1-\alpha)^{n}$ | $\frac{(1-\alpha)^{1-n}-1}{n-1}\quad if\ \ n>1$ | $\alpha(t)=1-\frac{1}{[1+[(n-1)K(t)]]^{\frac{1}{n-1}}}$ | Power ranking: induction stages Mampel 1940 |
| $f(\alpha)=(1-\alpha)^{-n}$ | $\frac{1}{n+1}\left([1-\alpha_o]^{n+1}-[1-\alpha]^{n+1}\right)$ | Parabolic law $\alpha(t)=\alpha_o+[(n+1)K(t)]^{\frac{1}{n+1}}$ for $0<K(t)<\frac{(1-\alpha_o)^{n+1}}{n+1}\approx\frac{1}{n+1}$ | One-dim. diffusion, D1 |
| n = random, m= random, p=0 $f(\alpha)=\alpha^{m}(1-\alpha)^{n}$ | Hypergeometric series | Impracticable | Arbitrary reaction Sestak-Berggren: SB – equation 1971 |
| $\alpha^{2/3}(1-\alpha)^{2/3}$ | $3\cdot x^{\frac{1}{3}}\cdot\text{hypergeom}\left[\begin{pmatrix}\frac{1}{3}\\1\end{pmatrix},\left(\frac{4}{3}\right),x\right]$ | Impracticable | Roginskii–Shultz 1928 Interface reactions (Kolmogorov) 1937 |
| $\alpha^{2/3}(1-\alpha)^{4/3}$ | $3\sqrt[3]{\frac{\alpha}{(1-\alpha)}}$ | $\alpha=\frac{1}{1+\left(\frac{3}{K(t)}\right)^{3}}$ | Modified Interface reactions (Kolmogorov) |
| $\alpha^{1-\frac{1}{n}}(1-\alpha)^{1+\frac{1}{n}}$ | $\frac{1}{n}\left[\frac{\alpha}{(1-\alpha)}\right]^{\frac{1}{n}}$ | $\alpha=\frac{1}{1+\left(\frac{n}{K(t)}\right)^{n}}$ | Modified logistic function |
| $(1-\alpha)[-\ln(1-\alpha)]^{1-\frac{1}{p}}$ | $[-\ln(1-\alpha)]^{\frac{1}{p}}$ | $\alpha=1-\exp\left(-(K(t))^{p}\right)$ | Nucleation-growth (KJMA equation) |
| n = (1-w), m=(1-q), p=0 $f(\alpha)=\alpha^{1-q}(1-\alpha)^{1-w}$ | Hypergeometric series | Impracticable | Early Jacobs and Tompkins relation Ng –conjecture 1975 |

Field Code Changed

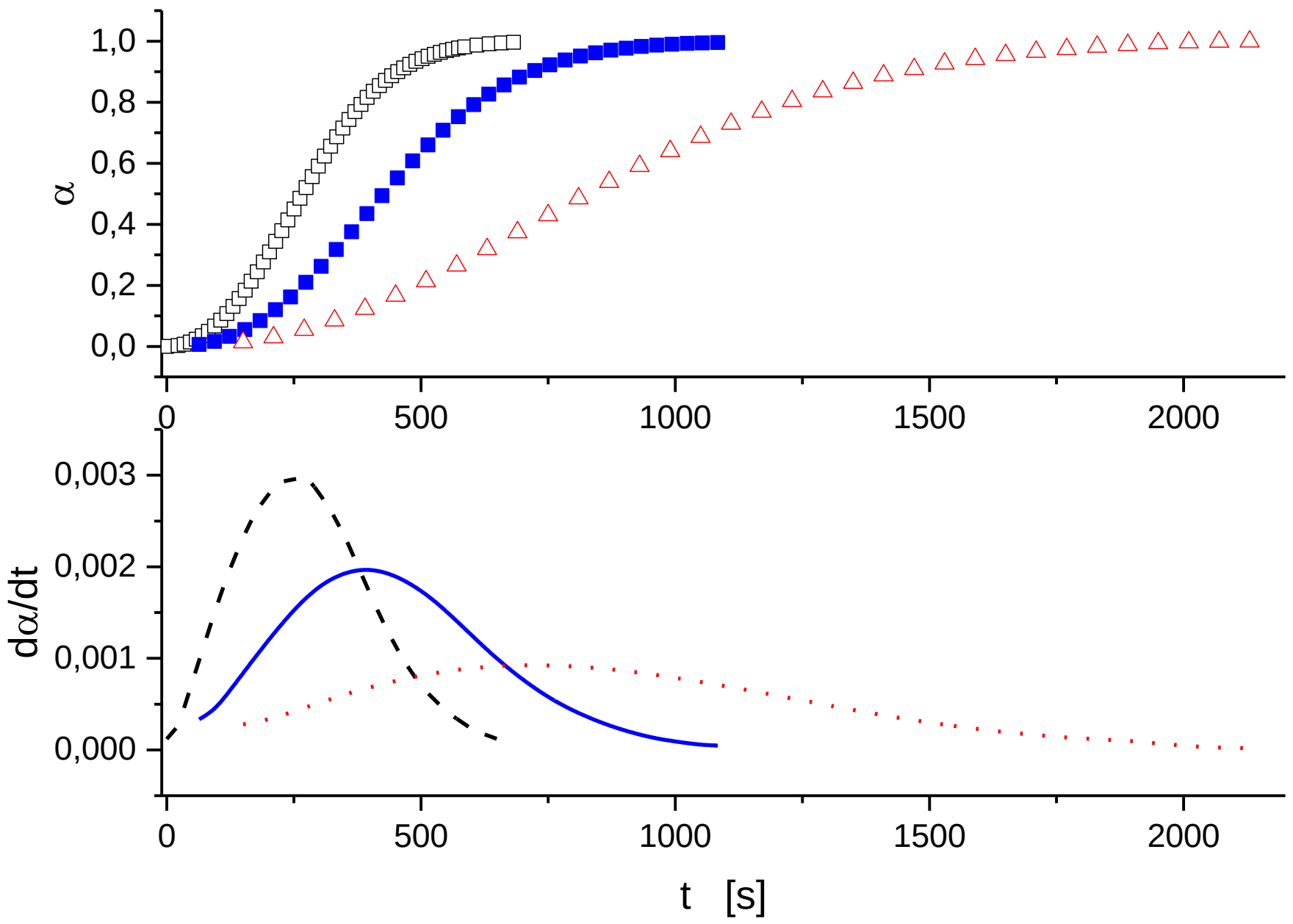

***Fig.1*** *The time, t, dependence degree of crystallization* $\alpha$ *(upper curve) and the overall crystallization rate* $\frac{d\alpha(t)}{dt}$ *of Isotactic Polypropylene. Data are from sample A in ref[41] . The open squares (□) and dashed line are for annealing temperature 125 C; the solid squares (■) and solid line are for 128 C and the open triangles (Δ) and the dotted line are for 130 C.*

**Mathematics behind the logistic functions**

The simplest assumption (*n=1, m=1, p=0*) [24-27,] listed in Table I, means that *f(α)* is proportional to the product of $\alpha$ and *(1 - α)*.

$$f(\alpha) = \alpha(1-\alpha) \qquad (7)$$

The term $\alpha$ accounts that the process is proceeding as a first-order reaction characterized by Prout and Tompkins [40] as autocatalytic or by Akulov [43] as autogenetic processes. In other words [24-27] the term, $\alpha$, is called *fertility* and the complementary term *(1 - α)* labeled *mortality* of a reactant yet ready to act in response of *(1 - α)*, *i.e.,* accounting that the studied process can proceed only in that part of the system that is not yet transformed. If Eq. (1) is under initial condition at *t=0*, there exists a starting fraction $\alpha_o$<<1 the integral $F(\alpha) \equiv \int_{\alpha_o}^{\alpha} \frac{d\alpha'}{f(\alpha')}$ which is becoming equal to

$$F(\alpha) = \ln\left[\frac{\alpha(1-\alpha_o)}{\alpha_o(1-\alpha)}\right]. \qquad (8)$$

For constant *k*, i.e. when Eq. (3) is valid, it is leading [15,24-27,36] to the classical logistic function of Verhulst [24]

$$\alpha = \frac{\alpha_o}{\alpha_o + (1-\alpha_o)e^{-kt}} \quad , \qquad (9)$$

As soon as $\alpha_o$<<1, Eq.(9) transforms in isothermal case to

$$\alpha = \frac{\alpha_o}{\alpha_o + e^{-kt}} \quad , \qquad (9a)$$

In the case of constant heating/cooling the rate q and in analogy with Eq. (9a) one obtains:

$$\alpha = \frac{\alpha_o}{\alpha_o + \frac{RT^2}{q\tau_o E}\exp\left(-\frac{E}{RT}\right)} \qquad (9b)$$

Also, from Eq.(3), it follows that a straight line is expected in coordinates $\ln\left[\frac{\alpha}{(1-\alpha)}\right]$ against time t having slope *k* and intersect $\ln\left[\frac{\alpha_o}{(1-\alpha_o)}\right]$. The degree of transformation $\alpha_p$, at which $\frac{d\alpha}{dt}$ has maximum, appears at $\alpha_p = 0.5$.

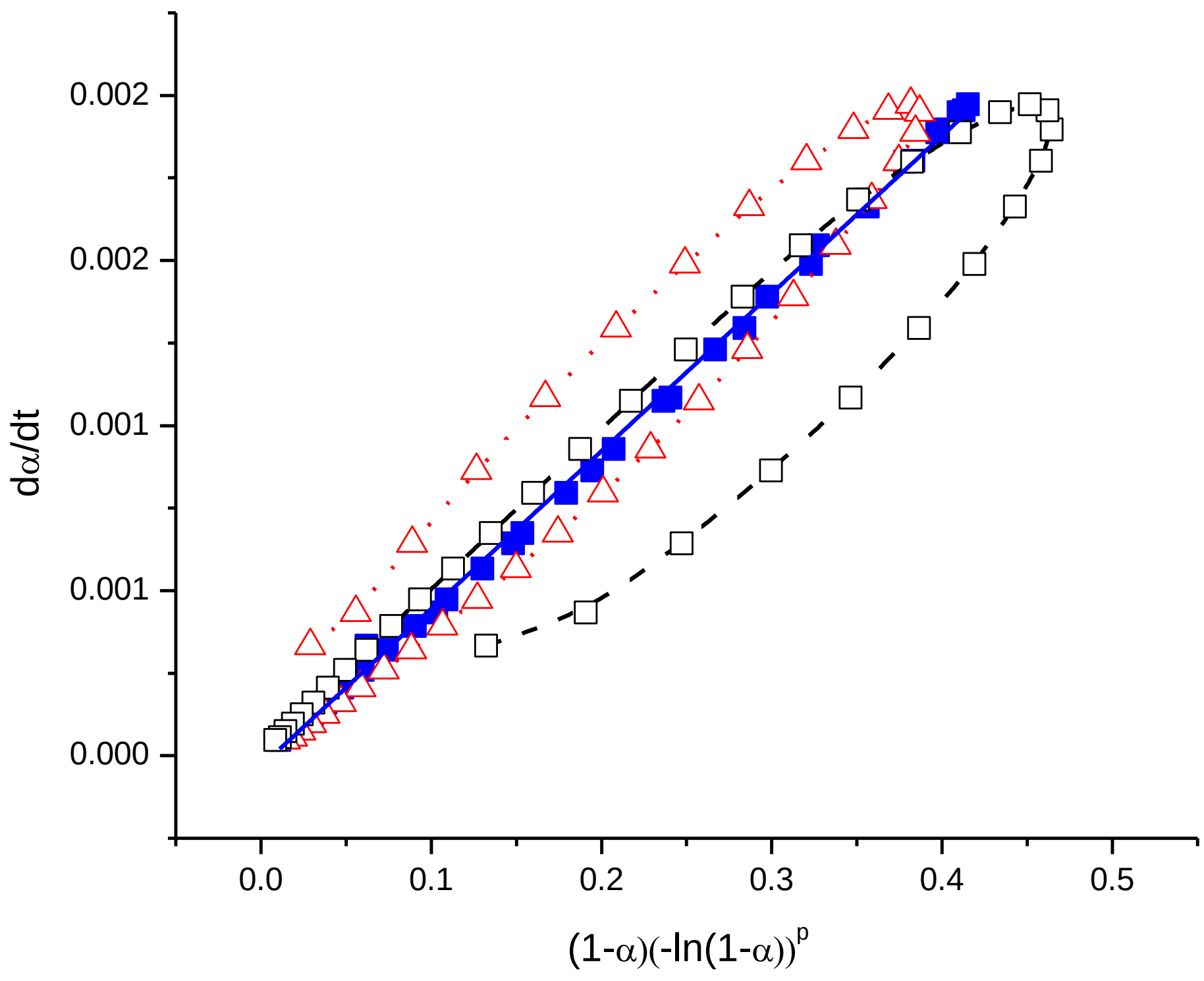

***Fig.2** Experimental data(from [41]) on kinetics of crystallization of Isotactic Polypropylene at 128 C in coordinates* $\frac{d\alpha(t)}{dt}$ *against* $(1-\alpha)(-\ln(1-\alpha))^p$ *for several values of p are as follows: the open triangles (Δ) are for Π =3.3 ; the open squares (□) are for Π =1.67. Only for Π =2.2 the data form a straight line, presented by the solid squares (■),*

Notwithstanding in Table I, the solutions are given within more generalized cases of $f(\alpha)$[9-13, 34–52] including non-integral or first choice fractal exponents [12,14,17], which may improvingly describe the prevalent cases of non-homogeneity and inherent geometrical heterogeneity in a relation to classical homogeneous reactions with a consistent concentration profile. It can be perceived in terms of the so called accommodation function [36,44] benefit to diagnostic limits of phenomenological kinetic models [6,11-13].

## Kinetics of overall crystallization

Overall crystallization is a complex process involving simultaneous nucleation and growth of separate crystallites. It was first described by Kolmogorov [38], Johnson & Mehl [39] and Avrami [1-3] and often abbreviated as *JMAK equation.* According to [1-3, 38-41] the degree α of overall phase transition of a system under constant external conditions (temperature and pressure) is presented by the equation

$$\alpha(t) = 1 - \exp\left(-\left(\frac{t}{\tau}\right)^{\Pi}\right) \qquad (10)$$

Althoughwidespread, there are reasons to doubtthe correctness ofthis formula. If Eq.(10) is correct, the rate of transformation in every moment should be given (see appendix 1) as

$$\frac{d\alpha(t)}{dt} = k(1-\alpha)(-\ln(1-\alpha))^{p} = \frac{\Pi}{t}(1-\alpha)(-\ln(1-\alpha)) \qquad (11)$$
$$where \quad p = 1 - \frac{1}{\Pi}; \quad k = \frac{\Pi}{\tau}$$

The best answer of the doubts is the comparison with existing experimental data. Fig.1 shows the time dependence of the degree of crystallization α (upper curve) and the overall crystallization rate

$\frac{d\alpha(t)}{dt}$ of Isotactic polypropylene annealed at 125C; 128C and at 130 C. Data are from sample *A* of ref. [41]. The values of pare objects of optimization until the plot of the data gives a straight line. A very sensitive test is to plot data in coordinates $\frac{d\alpha(t)}{dt}$ against $(1-\alpha)(-\ln(1-\alpha))^{p}$ , where the power is $p = 1 - \frac{1}{\Pi}$. If the value of *p* is properly determined a straight line with a slope *Π/τ* is formed. However, even minor errors in estimated *p* makes experimental data to form loops as shown in Fig. 2, where the investigational data[30] on kinetics of crystallization of Isotactic Polypropylene at 128 C are plotted in coordinates $\frac{d\alpha(t)}{dt}$ against $(1-\alpha)(-\ln(1-\alpha))^{p}$ for guessed values of *p*as follows: the red open triangles (Δ) are for *Π=3.3 (i.e. p=0.7)*; the open squares □ are for *Π =1.67 (i.e. p=0.4)*. In both cases the data lay along loops (but in opposite directions). Only for *Π =2.2 (i.e. p=0.55)* the data form straight line, presented by the solid squares (■).

In Fig.3 the experimental data for three annealing temperatures are given. For annealing temperatures 125 C and 128 C straight lines are formed for *Π =2.2* (i.e. *p =0.55*). For annealing temperature 130 C the straight line is formed if *Π=2 (i.e. p=0.5)*.

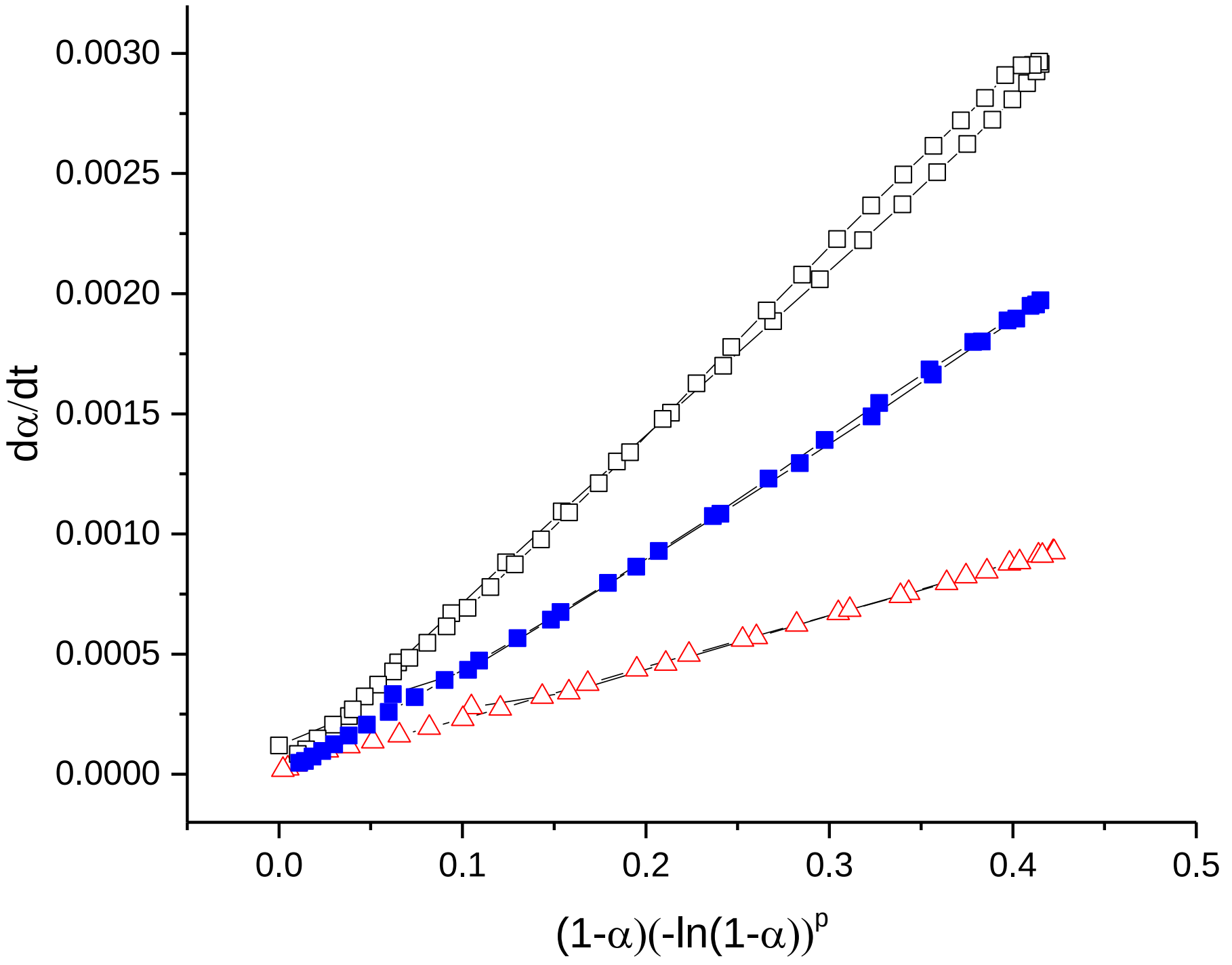

***Fig.3*** *Data from Fig.1 in coordinates in which, according to Eq.(11,) a straight line is expected with slopes □/τ. The value of Π is 2.2 for T=125 C (open triangles Δ) and 128 C (open squares □) and it is Π =2 for T=130 C (solid squares ■).*

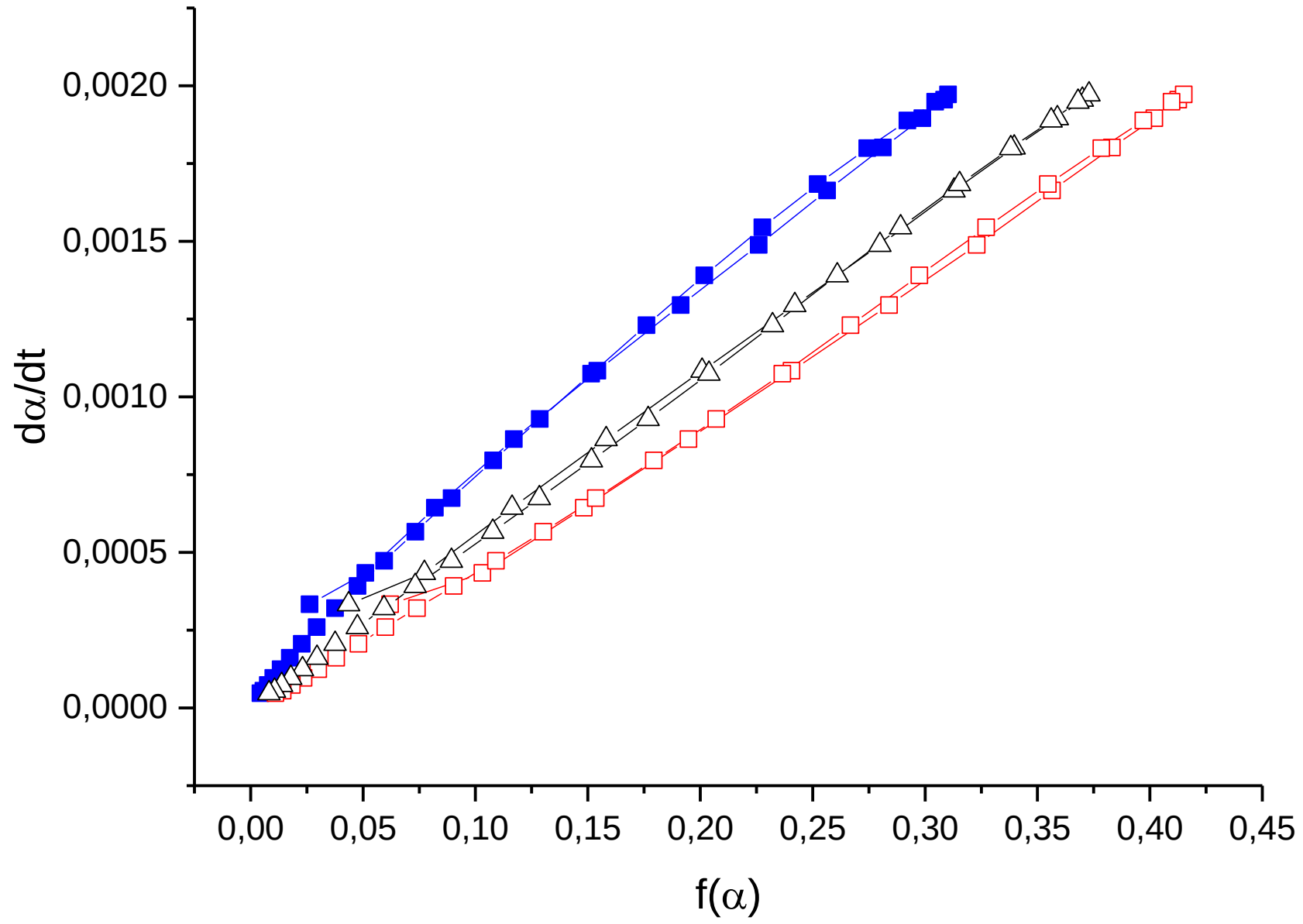

***Fig.4.*** *Data from Fig.1 for T=125 C in coordinates in which, according to Eq.(6) a straight line is expected with slopes p/τ. The solid squares (■) are for p =0 and m=0.72; the open triangles Δ are for m=0.25 and p=0.27 while the open squares (□) are for p=0.55 and m=0. In all cases n=1.*

**Discussion**

The ensuing Table I summarizes several possible forms of the $f(\alpha)$ functions, and the corresponding $F(\alpha)$ form of its integral elucidation. According to Eq.(2) the function $F(\alpha)$, can be solved for a limited number of combinations of values of *m*, *n* and *p*. However, the most of the reliableand consistent functions are solved [44] for the abridged form of $p=0$. The opposite case with $p\neq 0$ (i.e. under the power m considered $m=0$) is worth discussion for overall crystallization kinetics. Even for $p=0$ not all solutions of $F(\alpha)$ are of practical handling because the general solution leads to the undesired Hyper-Geometric-series. For that reason we discussed in Table I the particular cases for which the solution is of a practical use. The most popular assumption of *n=1* in Eq.(6) reflects the understanding that whatever happens, it takes place only in the part (1-$\alpha$) that is not yet transformed. If transformation process proceeds in *D*-dimensional space, the transformation rate is expected to be proportional to the size of periphery. This corresponds to an autocatalytic process with the fractional power $m = 1 - \frac{1}{D}$.

There are many recommendations about the coordinates in which to plot experimental data in order to extract information about the nature of the process and the values of the corresponding control parameters. Herewe offer afundamentally newmethod, especially reliable when *k(t)* does not change significantly with time. The idea is to plot the data in coordinates $\frac{d\alpha}{dt}$ against *f(α)*. If there are reasons to expect that *k(t)* does not change significantly with time,a straight line is expected, from the slope of itthe characteristic time of the process can be determined. An example of the method is presented in Figs.(2 and 3). If the values of the parameters *m, n* and *p* are not properly hosen, instead of straight lines the experimental data form loops, like the open points in Fig.2. Unfortunately, there are present too many adjustable parameters (*m, n* and *p*). Fig.4 thus illustrates that the same experimental data portray reasonably good straight lines for several sets of the *m, n* and *p* parameters. To simplify the problem, it is recommended to fix the parameter *n=1*. The reason is that the term *(1-α)* accounts for the fraction that still can undergo transformation. As for the other two parameters, the terms *α* and –*ln(1-α)* manifest similar behavior so that slight changes in the parameters *m* and *p* can compensate the tendencies and reasonable straight lines appear. It is a good idea to rely on reasonable physical model and to neglect one of the parameters on expense of the other. An example is the analyses of overall crystallization kinetics, which assumes that *m=0*. In the isothermal case and according to the *KJMA* equation (cf Table I) it is usually recommendable to plot data in the coordinates $\ln(-\ln(1-\alpha(t)))$ against ln t. Informative figures should be easily extracted from the expected straight line. The problem is that, for $\alpha \to 0$ as well as for $\alpha \to 1$, this approach is somehow misleading because in these cases $\ln(-\ln(1-\alpha))$ tends to go to infinity. So even minors errors in the determination accuracy of *α* turns to disastrous failure. For this reason (cf Refs. [54,55]), it is optional to better plot data in coordinates of α against *log(t)*. Although these coordinates sounds strange, there endures a trick that permits to extract information data safe while avoiding possible errors appearing in the previous coordinates.

The most difficult problem, still unsolved, is the case when there are enough reasons to expect that at least one of the parameters *m* or *p* are not constant. Example of this is the case of surface induced crystallization. At early stages a given number of active centers start to grow in three dimensions (*p=3*). Later on the growing crystals meet and form a hard core. The process continues inwards, in one direction so that $p<1$ and is changing continuously.

In conclusion let us remind that the design of a so-called *model of reaction mechanism* $f(\alpha)$ is only a part of a standard kinetic analysis of solid-state processes. The well quoted *SB*-equation (Eq. 7) can be likewise exploited in two different manners:

1/ As a resource of *individual mechanisms* (cf. Table I) providing the insight to standard mathematical-geometrical approach[4,12,13-22]. Some authors even tried to correlate its figures with the exponent p of *JMAK* model [42,36,46-48]. It reveals a gradual shifting of p along $0\rightarrow1.5\rightarrow2\rightarrow3$ producing the *m-n* pairs of *SB* exponents linked with the values of $0\text{-}1\rightarrow0.35\text{-}0.88\rightarrow0.54\text{-}0.83\rightarrow0.72\text{-}0.76$ [36,46-48]. Recently there appeared an attempt [48] to give the *m-n-p* correlation a mathematical figure such as $1/p = \ln(1-\alpha) - \ln(1-\alpha)\, n - \{(1-\alpha)/\alpha\} \ln(1-\alpha) m$ which is yielding upon a simplification the realtion $1/p \cong n - (3/4)\, m$. However, these attempts to associate the *SB* equation with a particular model of *JMAK* seems be somehow inappropriate because it is linking two in a way *incommensurable kinetics* exhibiting divergent philosophical strategies [53]: the geometrical *JMAK* versus the logistic *SB*.

2/ Therefore the SB approach can be seen as a *bridging equation* towards the novel method *model-free kinetics* [52, 56] which is trying to avoid mathematical idealization in construction models based on idealized geometrical bodies.

The latter stays away from the use of somehow ´irresolute´ mathematical form of an exponential, Eq. (4), within its crucially reciprocal linkage between the exponential term $(E/T)$ and its pre-exponential factor $(1/\tau_o)$ [49-51] which is a mathematically irremovable correlation [49].

In addition we have to take care about some yet unsettled impacts and yet unresolved links which are interfering in the so far customary methods of kinetic evaluations and which nonetheless are not solved critically enough [53]. Among others we may indicate the effect of the proximity to equilibrium [57,58] or a generalized impact of both the heat inertia [59,60] and temperature gradients [59,60] always existing within the bulk of a sample under thermal study at nonzero heating rates.

**Appendix 1**

From Eq.(10) one has

$$\exp\left(-\left(\frac{t}{\tau}\right)^{\Pi}\right)=1-\alpha\,;\quad \left(\frac{t}{\tau}\right)^{\Pi}=-\ln(1-\alpha)\;;t=-\tau\ln(1-\alpha)^{\frac{1}{\Pi}} \qquad (1a)$$

The derivative of Eq (10) is

$$\frac{d\alpha(t)}{dt}=\frac{\Pi}{t}\left(\frac{t}{\tau}\right)^{\Pi}\exp\left(-\left(\frac{t}{\tau}\right)^{\Pi}\right) \qquad (2a)$$

So that the combination of Eqs (1a and 2a) leads to

$$\frac{d\alpha(t)}{dt}=\frac{\Pi}{\tau}(1-\alpha)\left(-\ln(1-\alpha)^{1-\frac{1}{\Pi}}\right)=\frac{\Pi}{t}(1-\alpha)(-\ln(1-\alpha)) \qquad (3a)$$

While the term $(1-\alpha)$ accounts for the role of the volume that is still unoccupied by the new phase, the logarithmic term sounds strange. Originally the *KJMA* equation was derived from the integral form of the extended degree of transformation $\alpha_{ext}$ , i.e. the fraction of the transformed regions under condition they can overlap and the assumption $\mathrm{d}\alpha(\mathrm{t})=(1\text{-}\alpha)\mathrm{d}\alpha_{\mathrm{ext}}$ that at any moment the really transformed part $\frac{\mathrm{d}\alpha}{dt}$ is proportional to the product of $\frac{\mathrm{d}\alpha_{ext}}{dt}$ and the part $(1\text{-}\alpha)$ that is available for transformation; i.e.

$$\frac{\mathrm{d}\alpha(\mathrm{t})}{dt}=(1\text{-}\alpha)\frac{\mathrm{d}\alpha_{\mathrm{ext}}}{dt}\,. \qquad (4a)$$

The comparison of Eqs (3a and 4a) shows that

$$\frac{d\alpha_{ext}(t)}{dt}=\frac{\Pi}{\tau}\left(-\ln(1-\alpha)^{1-\frac{1}{\Pi}}\right)=\frac{\Pi}{t}(-\ln(1-\alpha)) \qquad (5a)$$

To give a good physical reason of Eq.(5a) is rather difficult, so this offers reason to use also other logistic functions.

**Acknowledgement:**

The results were developed within the CENTEM project, reg. no. CZ.1.05/2.1.00/03.0088 that is co-funded from the ERDF as a part of the MEYS - Ministry of Education, Youth and Sports OP RDI Program and, in the follow-up sustainability stage supported through the CENTEM PLUS (LO 1402) by financial of above MEYS under the "National Sustainability Program I" and developed under the Joint Research Laboratory between the New Technologies Centre of the University of West Bohemia in Pilzen and the Section of Solid-State Physics of Institute of Physics, AcSc in Prague. Deep thanks are due to the cooperative discussions with P. Holba (West Bohemian University in Pislen), N. Koga (Hiroshima University), J. Málek (University of Pardubice), J.J. Mareš (Institute of Physics in Prague), C. Russel (Friedrich-Schiller University in Jena) and P. Šimon (Slovak Technical University in Bratislava).

## References

1. Avrami, M., *J. Chem. Phys.*, *7*, 1103 (1939)
2. Avrami, M., *J. Chem. Phys.*, *8*, 212 (1940)
3. Avrami, M., *J. Chem. Phys.*, *9*, 177 (1941)
4. Hulbert H.F. *J. Br. Ceram. Soc.* 6, 11 (1969)
5. Málek, J.; Criado, J. *Thermochim. Acta. 203*, 25 (1992)
6. Šesták J, Málek J. *Solid State Ionics* 63/65, 254 (1993)
7. Málek J. *Thermochim. Acta.*, 267, 61 (1995)
8. Koga N. *J Therm Anal.* 49,45 (1997)
9. Mamleev V, Bourbigot S, LeBras M, Duquesne S, Šesták J. *Phys Chem Chem Phys.*, 2, 4796 (2000).
10. Šesták J. *J Thermal Anal Calor.*, 110, 5 (2012).
11. Koga N, Šimon P, Šesták J. Some fundamental and historical aspects of phenomenological kinetics in solid-state studied by thermal analysis, chap. 1. In: *Thermal analysis of micro-, nano- and non-crystalline materials*, (Šesták J, Šimon P, editors). Berlin: Springer; 2013. p. 1–28.
12. Šesták J. Mechanism and kinetics of heterogeneous non-catalyzed reactions, Chapt. 8 in his book: *Thermophysical Properties of Solids: their measurements and theoretical thermal analysis*, 172-208, Elsevier, Amsterdam 1984 and in Russian translation: *´Teoretičeskij termičeskij analyz´*, pp. 186-222, Mir, Moscow 1987
13. Khawam A, Flanagan D.R.*J. Phys. Chem. B*,110, 17315 (2006)
14. Gotor, F. J.; Criado, J. M.; Málek, J.; Koga, N., *J. Phys. Chem. A, 104*, 10777 (2000).
15. Burnham A.K., Weese R.K., Weeks B.L.,*J. Phys. Chem. B.*, 108, 19432 (2004).

16. Perejón, A.; Sánchez-Jiménez, P. E.; Criado, J. M.; Pérez-Maqueda, L. A., *J. Phys. Chem. B*, *115*, 1780 (2011).

17. Cai, J.; Liu, R., Weibull. *J. Phys. Chem. B* , *111*, 10681 (2007).

18. Wada, T.; Koga, N.,*J. Phys. Chem. A*, *117*, 1880 (2013).

19. Koga, N.; Yamada, S.; Kimura, T., *J. Phys. Chem. C*, *117*, 326 (2013)

20. Noda, Y.; Koga, N., *J. Phys. Chem. C* , *118*, 5424 (2014).

21. Yoshikawa, M.; Yamada, S.; Koga, N.,*J. Phys. Chem. C* 2014, 118, 8059 (2014).

22. Kitabayashi, S.; Koga, N.,*J. Phys. Chem. C*, *118*, 17847 (2014).

23. R. Svoboda, J. Málek. *J. Chem. Phys.*, 141, 224507 (2014).

24. Verhulst P.F. Notice sur la loi que la population poursuit dans son accroissement. Correspondance mathematique et physique 1838; X: 113–121.

25. Verhulst P.F. Mem. Acad. R. Bruxelles,18,1 (1844).

26. West B, Rrigolini P.G. Complex webs. Cambridge: Cambridge University Press; 2011.

27. Avramov I.. *Phys A.*, 379, 615 (2007).

28. White P.S. *Bot Rev*., 45, 229 (1979).

29. Mooney H.A, Godron M, editors. Disturbance and ecosystems. New York: Springer; 1983.

30. Romme W.H, Knight D.H. Landscape diversity: the concept applied to Yellowstone Park. *Bioscience.* 1982; 32:664.

31. Pickett S.T.A, White P.S, editors. Ecology of natural disturbance and patch dynamics. New York: Academic Press; 1985.

32. Turner M.G, editor. Landscape heterogeneity and disturbance. New York: Springer; 1987.

33. Young D.A. Decomposition of solids. Oxford: Pergamon Press; 1966.

34. Šesták J. Modeling of reaction mechanism: use of Euclidian and fractal geometry, Chapt. 10. In his book: *Science of heat and thermos physical studies: a generalized approach to thermal analysis*. pp. 276-314, Elsevier, Amsterdam 2005.

35. Šimon P. . *Thermochim. Acta* 520, 156 (2011)

36. Málek J, Criado JM. *Thermochim Acta.* 6: 179 (1973)

37. I. Avramov, J. Šesták, *J Therm Anal Calorim* , 118, 1715 (2014)

38. Kolmogorov. *Izv.* Akad. Nauk SSSR. *Ser. Matem*. (in Russian) 1,355 (1937)

39. Johnson W., Mehl R. *Trans. AIME*, 135, 416 (1939)

40. Prout E.G, Tompkins F.C. *Trans Faraday Soc*., 40, 488 (1944).

41. Celli, A., Zanotto E, Avramov I, J. *Macromol. Sci. Part B—Physics* 42, 387 (2003)

42. Illekova E, Šesták J. Crystallization kinetics of metallic micronano- and non-crystalline glasses, chap. 13. In:.*Thermal analysis of micro-, nano- and non-crystalline materials.* Šesták J, Simon P, editors, Berlin: Springer; 2013. p. 257–90.

43. Akulov N.S, (in Russian); *Compt. Rend. Acad. Sci. USSR*, 27, 135-138 (1940) (English translation)

44. Šesták J. *J Therm Anal.* 410,36 (1997).

45. Bagchi T.P, Sen P.K. *Thermochim Acta.* 51,175 (1981).

46. J. Málek, T. Mitsuhashi. *J. Am. Ceram. Soc*.,83, 2103 (2000)

47. Málek J. *Thermochim. Acta* 267: 61 (1995)

48. Arshad, M.A, Maaroufi, A. *J Non-Cryst. Solids*, 413: 53 (2015)

49. Koga N, Šesták J, Málek J. *Thermochim Acta*. 188,333 (1991).

50. Koga N. *Thermochim. Acta* 244: 1 (1994)

51. Galwey AK, Mortimer M. *Int. J. Chem. Kinet.* 38: 464 (2006)

52. Šimon P. *J. Therm. Anal. Calorim*. 79, 703 (2005).

53. Šesták J, *Thermochim Acta* 611, 26 (2015)

54. Avramov I, Avramova K, Russel C *J. Cryst. Growth* 285 394 (2005)

55. Avramov I, Avramova K, Russel C, *J. Non-Cryst. Sol* 356, 1201 (2010)

56. Šimon P, Dubaj T, Cibulková Z. *J Therm Anal Calorim,* 120, 231 (2015)

57. Holba P. Šesták J. *Zeit. physik. Chem. N.F.* 80: 1 (1972).

58. P. Holba, *J Thermal Anal Calorim*, 120: 175 (2015)

59. Holba P. Šesták J *J Thermal Anal Calorim,* 113, 1633 (2013)

60. Holba P. Šesták J. Sedmidubský D. Heat transfer and phase transition at DTA experiments. Chapter 5 in book:*Thermal Analysis of Micro-, Nano- and Non-Crystalline Materials* (J. Šesták, P. Šimon. Editors), pp. 99-134, Springer Berlin 2013